\documentclass[epj]{svjour}
\usepackage{graphics}
\usepackage{amssymb}
\newcommand{\be}{\begin{equation}}
\newcommand{\ee}{\end{equation}}
\newcommand{\bd}{\begin{displaymath}}
\newcommand{\ed}{\end{displaymath}}
\newcommand{\BE}{\begin{eqnarray}}
\newcommand{\EE}{\end{eqnarray}}

\newcommand{\avg}[1]{\left\langle{#1}\right\rangle}
\begin{document}
\title{Relative population size, co-operation pressure and strategy correlation in two-population evolutionary dynamics}
\titlerunning{Two-population evolutionary dynamics}
\author{Tobias Galla}
%
%
\institute{The University of Manchester, School of Physics and Astronomy, Schuster Building, Manchester M13 9PL, United Kingdom\\
The Abdus Salam International Centre for Theoretical Physics, Strada Costiera 11, 34014 Trieste, Italy}
\date{\today}
%
\abstract{We study the coupled dynamics of two populations of random replicators by means of statistical mechanics methods, and focus on the effects of relative population size, strategy correlations and heterogeneities in the respective co-operation pressures. To this end we generalise existing path-integral approaches to replicator systems with random asymmetric couplings. This technique allows one to formulate an effective dynamical theory, which is exact in the thermodynamic limit and which can be solve for persistent order parameters in a fixed-point regime regardless of the symmetry of the interactions. The onset of instability can be determined self-consistently. We calculate quantities such as the diversity of the respective populations and their fitnesses in the stationary state, and compare results with data from a numerical integration of the replicator equations.}

%
\maketitle
\section{Introduction}
The evolutionary dynamics of populations of interacting species is
often described by so-called replicator equations (RE)
\cite{Book4,Book5}. Each species here carries a fitness, and species
who are fitter than the average increase in concentration whereas the
weight of species less fit than average decreases. RE have also found
applications in game theory and economics \cite{Book6}, and an
equivalence to Lotka-Volterra equations of population dynamics can be
established. In the context of game theory RE describe situations in
which a game is played repeatedly by players chosen randomly from
populations of agents. Each individual agent here plays one fixed
(so-called pure) strategy throughout his lifetime, and reproduction
occurs in proportion to the payoff accrued. Offspring then inherit the
strategy of their parent. In this situation the replicators are hence
the pure strategies of the game under consideration. In this paper we
will use the interpretations of RE in population dynamics and in
evolutionary game theory simultaneously, and will hence refer to the
replicating entities as `species' and `strategies'
synonymously. Similarly, `fitness' and `payoff' will be used
interchangeably.

The analysis of eco-systems and replicators with fixed random
couplings was initiated by the seminal work of \cite{May}, and
although the study of real-world eco-systems has since then moved on
the incorporate more realistic and evolving networks (see
e.g. \cite{McKane_Drossel_2005} and references therein), replicator
systems with quenched random couplings are still of interest from the
statistical mechanics perspective and as an analytically tractable
benchmark. Techniques from disordered systems theory have first been
applied to such models in
\cite{DO,OD,OD2,ParisiBiscari} and focus both on static approaches
based on the replica method and on dynamical studies by means of
generating functional techniques. The latter approach here has the
advantage of being able to address couplings of general symmetry
\cite{OD,galla}, whereas replica theory is restricted to cases of
symmetric interaction (see \cite{OF1,OF2,OF3,OL,SF,Tokita} for further
replica studies of random replicator models). In the course of these
studies complex non-trivial behaviour of random replicator systems has
been identified, with several types of phase transitions between
stable and unstable regimes, and multiple patterns of ergodicity
breaking. While most of the existing studies have focused on
so-called single-population models, corresponding to games with only
one type of player, multi-population models have recently been
addressed in \cite{epl}, where a relation between the stability of
two-population replicator systems and the corresponding two-player
bi-matrix games has been considered. These are games in which the
two interacting players take different roles (e.g. female versus
male), and may hence have different sets of strategies at their disposal (as opposed to single-population games such as rock-papers-scissors, where the strategy pool is identical for both players). Static studies of random matrix games can also be found in \cite{berg,berg2,berg3}.

The aim of the present work is to extend the studies of \cite{epl} to
the case of two interacting populations of different relative size and
to incorporate correlations between payoff matrices as well as
heterogeneity in the co-operation pressures of species. We first present the details of the model in the next section, and then formulate an effective macroscopic theory in the thermodynamic limit. Specific results are then reported in section \ref{sec:results}, where we test the predictions of the resulting fixed-point theory against numerical simulations. A summary and outlook conclude the paper.

\section{Model}
We consider two populations of replicators (abbreviated as $P1$ and
$P2$ in the following), where we denote the species in system $P1$ by
$i=1,\dots,N_1$, and the ones in population $P2$ by
$j=1,\dots,N_2$. $N_1$ and $N_2$ are hence the sizes of the two
populations (equivalently the number of pure strategies available to
the two types of players of the bi-matrix game). We will consider the
general case $N_1\neq N_2$, and will write $N_1=\alpha N$ and
$N_2=\alpha_2 N$ in the following. The statistical mechanics analysis
is then concerned with the limit $N,N_1,N_2\to\infty$, where the
ratios $\alpha_1=N_1/N$ and $\alpha_2=N_2/N$ remain finite. The aspect
ratio $\alpha=\alpha_1/\alpha_2$ is hence a control parameter of the
model.

We will refer to the species in $P1$ as species of type $X$, and to species in $P2$ as species of type $Y$. The composition of $P1$ at time $t$ is then characterised by a concentration vector $(x_1(t),\dots,x_{N_1}(t))$, where $x_i$ indicates the relative weight of species $i$ in $P1$. Similarly, the configuration of $P2$ can be described by $(y_1(t),\dots,y_{N_2}(t))$. In the context of evolutionary game theory, vectors of this type characterise mixed strategies \cite{Book6}, with e.g. $y_j$ being proportional to the probability of a player of type $Y$ using pure strategy $j\in\{1,\dots,N_2\}$. We will in the following use the normalisation $N_1^{-1}\sum_i x_i=1$ and $N_2^{-1}\sum_j y_j=1$, which will allow us to formulate a well-defined thermodynamic limit of the problem.

Species of type $X$ are assumed to interact only with species of type $Y$ and vice versa, and the populations are taken to follow the following replicator equations
\BE
\dot x_i&=&x_i\left(-2u_1 x_i+\sum_{j=1}^{N_1} a_{ij} y_j-f_1\right), \nonumber \\
\dot y_j&=&y_j\left(-2u_2 y_j+\sum_{i=1}^{N_2} b_{ji} x_i-f_2\right).\label{eq:re}
\EE
$u_1$ and $u_2$ are here the co-operation pressures acting on the two populations. Depending on their strengths $u_1$ and $u_2$ limit the growth of individual species and drive $P1$ and $P2$ towards states with many surviving species and high diversity \cite{Book5}. We will mostly consider the situation in which $u_1,u_2\geq 0$. The payoff matrices (or inter-species couplings) $\{a_{ij},b_{ji}\}$ are drawn from a Gaussian distribution with the following first two moments:
\BE
&\overline{a_{ij}}=\overline{b_{ji}}=0, \nonumber \\
&\overline{a_{ij}^2}=\overline{b_{ji}^2}=\frac{1}{N}, \nonumber \\
&\overline{a_{ij}b_{ji}}=\frac{\Gamma}{N}.
\EE
The overbar denotes an average over the disorder, i.e. over samples of the random payoff matrices. All other covariances are taken to vanish. The scaling of the covariance matrix with the system size $N$ is here
again chosen to guarantee a well-defined thermodynamic limit
$N\to\infty$, with which the statistical mechanics theory will be
concerned. $\Gamma\in[-1,1]$ is a model parameter, characterising the
symmetry or otherwise of the interaction: for $\Gamma=-1$ we have
$a_{ij}=-b_{ji}$ corresponding to prey-predator relations in
population dynamics, and to a zero-sum game in the context of
evolutionary game theory. For $\Gamma=0$ $a_{ij}$ and $b_{ji}$ are
uncorrelated, while $\Gamma=1$ corresponds to the case of symmetric
interaction, $a_{ij}=b_{ji}$. Intermediate values of $\Gamma$ allow
for a smooth interpolation between these cases. $f_1$ and $f_2$
finally are the time-dependent mean fitnesses of species in the two respective
populations,
\BE
f_1&=&N_1^{-1}\sum_{i=1}^{N_1}x_i f^{(1)}_i \nonumber \\
f_2&=&N_2^{-1}\sum_{j=1}^{N_2}y_j f^{(2)}_j,
\EE
where $f_i^{(1)}=-2u_1x_i+\sum_j a_{ij}y_j$ denotes the fitness of species $i$ (the superscript indicates that this species belongs to $P1$). $f_j^{(2)}$ is defined similarly. These definitions ensure that the replicator equations (\ref{eq:re}) conserve the normalisations $N_1^{-1}\sum_{i=1}^{N_1}x_i=N_2^{-1}\sum_j y_j=1$ in time. Initial conditions in our simulations are chosen to respect this normalisation.

\section{Statistical mechanics theory}
\subsection{Uncorrelated strategies}
The dynamics of the model can be reduced to two coupled stochastic
processes for a representative pair of `effective' species, one from each
population. We will not detail the mathematical steps here, but will
only quote the final outcome. Similar calculations can be found in
\cite{OD,galla} or, in a different context, in the textbook
\cite{Book2}. The generating functional analysis delivers the following effective dynamics:
\BE
\dot x(t) &=& x(t)\bigg[-2u_1 x(t) +\Gamma\alpha_2\int dt' G_2(t,t')x(t')+\eta_1(t)-f_1(t)\bigg],\nonumber \\
\dot y(t) &=& y(t)\bigg[-2u_2 y(t) +\Gamma\alpha_1\int dt' G_1(t,t')y(t')+\eta_2(t)-f_2(t)\bigg],
\nonumber \\\label{eq:effproc}
\EE
where $\eta_1(t)$ and $\eta_2(t)$ represent coloured Gaussian noise of zero mean and with covariances to be determined self-consistently as
\BE
\avg{\eta_1(t)\eta_1(t')}&=&\alpha_2\avg{y(t)y(t')},\nonumber \\
\avg{\eta_2(t)\eta_2(t')}&=&\alpha_1\avg{x(t)x(t')}, \nonumber \\
\avg{\eta_1(t)\eta_2(t')}&=&0.
\EE
$\avg{\cdot}$ here refers to an average over realisations of the effective processes, i.e. over paths of ${\eta_1(t),\eta_2(t)}$. The response functions $G_1(t,t')$ and $G_2(t,t')$, in turn, are given by
\BE
G_1(t,t')&=&\avg{\frac{\delta x(t)}{\delta \eta_1(t')}}, \nonumber \\
G_2(t,t')&=&\avg{\frac{\delta y(t)}{\delta \eta_2(t')}}
\EE
respectively. As usual in the dynamics of disordered systems these processes are non-Markovian (as reflected by the retarded interaction terms), and the resulting single-particle noises $\{\eta_1(t),\eta_2(t)\}$ exhibit non-trivial temporal correlations. The description in terms of the above effective processes is equivalent to the original problem in the thermodynamic limit in the sense that combined disorder-species-averages in the microscopic problem can be evaluated as averages over realisations of the effective single-particle noise (see \cite{Book2} for further technical details regarding the generating functional technique).

Following \cite{OD}, we proceed with a fixed-point analysis of (\ref{eq:effproc}), based on the assumption of time-translation invariance and finite integrated responses
\be
\chi_n=\lim_{t\to\infty} \int_0^\infty d\tau G_n(t+\tau,t)<\infty
\ee 
($n=1,2$). We also write $q_1=\avg{x^2}$ and $q_2=\avg{y^2}$, where $x$ and $y$ denote the fixed point values of the single-species concentrations obtained from (\ref{eq:effproc}) as $t\to\infty$. Since each realisation of the single-species trajectories is a time-dependent stochastic process, the resulting fixed point values $x$ and $y$ are static random variables. See \cite{OD,galla} for further technical details regarding this approach.

The fixed-point ansatz then leads to the following six equations for the persistent order parameters \\ $\{q_1,q_2,\chi_1,\chi_2,f_1,f_2\}$:
\BE
\chi_1(2u_1-\alpha_2\Gamma\chi_2)&=&g_0(\Delta_1),\nonumber \\
\chi_2(2u_2-\alpha_1\Gamma\chi_1)&=&g_0(\Delta_2),\nonumber \\
(\alpha_2q_2)^{-1/2}(2u_1-\alpha_2\Gamma\chi_2)&=&g_1(\Delta_1), \nonumber \\
(\alpha_1q_1)^{-1/2}(2u_2-\alpha_1\Gamma\chi_1)&=&g_1(\Delta_2), \nonumber \\
q_1/(\alpha_2 q_2)(2u_1-\alpha_2\Gamma\chi_2)^2&=&g_2(\Delta_1), \nonumber \\
q_2/(\alpha_1 q_1)(2u_2-\alpha_1\Gamma\chi_1)^2&=&g_2(\Delta_2), 
\label{eq:fp}
\EE
where $\Delta_1=-f_1/\sqrt{\alpha_2q_2}$ and $\Delta_2=-f_2/\sqrt{\alpha_1q_1}$ and where $g_k(\Delta)=\int_{-\infty}^\Delta \frac{dz}{\sqrt{2\pi}}e^{-z^2/2}(\Delta-z)^k$ for $k\in\{0,1,2\}$. These equations are readily solved to yield the persistent order parameters $\{q_1,q_2,\chi_1,\chi_2,f_1,f_2\}$ as functions of the model parameters $\{u_1,u_2,\alpha_1,\alpha_2,\Gamma\}$. It is here efficient to proceed as proposed in \cite{SF} and to obtain parametric solutions in terms of $\{\Delta_1,\Delta_2\}$, i.e. to fix the values of these two quantities, and then to obtain $\{q_1,q_2,\chi_1,\chi_2,u_1,u_2\}$ from the left-hand sides of (\ref{eq:fp}). Note that $\phi_1\equiv g_0(\Delta_1)$ and $\phi_2\equiv g_0(\Delta_2)$ are the fractions of surviving species, in the two respective populations. In the context of game theory these expressions correspond to the fractions of pure strategies played with non-zero probabilities \cite{berg,berg2,berg3}. As seen in \cite{galla} $1/q_1$ and $1/q_2$ serve as measures of the resulting diversities as well. $q_1$ is here given by $q_1=N_1^{-1}\sum_i \overline{x_i^2}$ at the fixed point, and a similar definition of $q_2$ applies. If e.g. all species are present at equal concentration $x_i=1$ in $P1$ ($i=1,\dots,N_1)$, then $1/q_1=1$. If however only a few species survive in $P_1$, then $1/q_1\to 0$ (in the extreme case of only one species surviving, say $x_1=N_1$ and $x_i=0$ for $i>1$ one has $q_1=N_1$ which tends to infinity in the thermodynamic limit). A detailed inspection shows that $1/q_1$ and $1/q_2$ are closely related to what is referred to Simpson's index of diversity in ecology \cite{simpson}.
\subsection{Correlated strategies}
We will below also consider the case of correlated strategies as proposed in \cite{berg,berg2}. Here, the moments of the matrix elements $\{a_{ij},b_{ji}\}$ are further constrained by imposing
\BE
\overline{a_{ij}a_{kj}}=\frac{c_1}{N}\frac{1}{N}, \nonumber \\
\overline{a_{ik}a_{jk}}=\frac{c_2}{N}\frac{1}{N},
\EE
with correlation parameters $c_1,c_2\geq 0$. We will furthermore restrict the analysis to the case $\Gamma=-1$ when considering strategy correlations, we then have $b_{ji}=-a_{ij}$. $c_1$ and $c_2$ measure the correlations within rows and columns of the payoff matrix. I.e. if $c_2$ is large, the elements \\$a_{1j},a_{2j},a_{3j},\dots, a_{N_1 j}$ bear some correlation for any fixed $j$, reflecting a reduction of the variability of strategies in $P1$ (all pure strategies $i=1,2,3,\dots,N_1$ in $P1$ bear some degree of similarity). Similarly, an increased value of $c_1$ makes pure strategies in $P2$ more alike, and hence reduces the strategic options of individuals in that population. 

These strategy correlations can be incorporated in the path-integral
analysis, and lead to the following modifications in the resulting
effective processes:
\BE
\dot x &=& x\bigg[-2u_1 x -\int dt' G_2(t,t')(x(t')+c_2)+\eta_1(t)-f_1(t)\bigg],\nonumber \\
\dot y &=& y\bigg[-2u_2 y -\int dt' G_1(t,t')(y(t')+c_1)+\eta_2(t)-f_2(t)\bigg].\nonumber \\
\EE
The covariances of the effective single-species noises are now given by
\BE
\avg{\eta_1(t)\eta_1(t')}&=&\avg{y(t)y(t')}+c_1,\nonumber \\
\avg{\eta_2(t)\eta_2(t')}&=&\avg{x(t)x(t')}+c_2,
\EE
and $\avg{\eta_1(t)\eta_2(t')}=0$ as before. The resulting equations for the persistent order parameters undergo the corresponding changes as well (we do not report them here).

\subsection{Stability}
The stability of the fixed point identified and used in the above ansatz can be investigated by means of a linear expansion about the assumed fixed point values of the concentrations of the effective pure strategy frequencies $x$ and $y$ and of the stationary noise variables $\eta_1$ and $\eta_2$. We will not go into details here, as the analysis follows the lines of \cite{OD}. The final outcome is that the system is stable whenever
\be
\frac{\phi_1}{\chi_1^2}\frac{\phi_2}{\chi_2^2}>\alpha_1\alpha_2,
\ee
and that non-trivial and non-decaying fluctuations may arise whenever this condition is violated, so that the above fixed-point ansatz breaks down and Eqs. (\ref{eq:fp}) can no longer be expected to describe the behaviour of the model accurately. For $\alpha_1=\alpha_2=1$ the above condition reduces to the one reported in \cite{epl}. We also remark that, if additionally $u_1=u_2\equiv u$, Eqs. (\ref{eq:fp}) reduce to the one-population results of \cite{OD,galla}. The onset of instability then occurs at $u_c=(1+\Gamma)/(2\sqrt{2})$, as first reported in \cite{OD}.  
\section{Results}\label{sec:results}
\begin{figure}\vspace{0.5cm}
\resizebox{0.38\textwidth}{!}{%
  \includegraphics{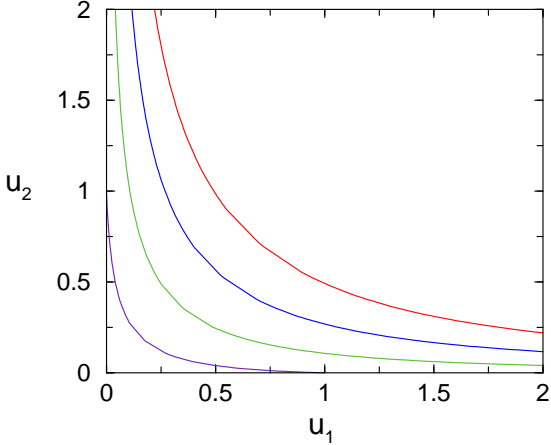}
}
\vspace*{4mm} \caption{(Colour on-line) Phase diagram of the model with two co-operation pressures $u_1$ and $u_2$ ($\alpha_1=\alpha_2=1, c_1=c_2=0$). The lines show the onset of instability, with the stable phase above the respective lines. Phase boundaries are shown for $\Gamma=1,0.5,0,-0.5$ from top to bottom.}
\label{fig:pg}
\end{figure}

\begin{figure}[t]
\vspace{0.75cm}
\resizebox{0.38\textwidth}{!}{%
  \includegraphics{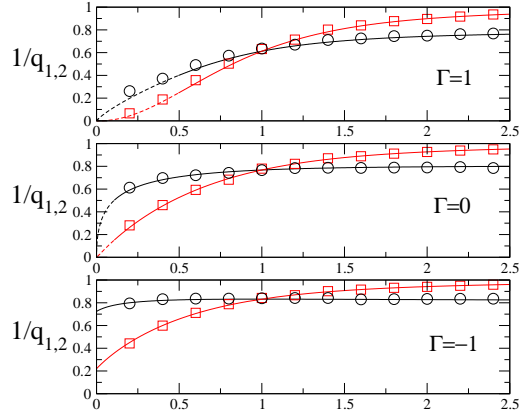}
}
\vspace*{4mm} \caption{(Colour on-line) Diversities $1/q_1$ and $1/q_2$ of the two populations as a function of the co-operation pressure in population $2$. $u_1=1$ fixed  ($\alpha_1=\alpha_2=1, c_1=c_2=0$. Symbols are from simulations for $N=100$ ($5000$ integration steps, $10$ samples), with black circles corresponding to $1/q_1$ and red squares to $1/q_2$. $\Gamma=1,0,-1$ in the top, middle and bottom panel respectively. Lines are from theory (solid in stable phase, dashed in unstable regime).}
\label{fig:u1u2}
\end{figure}

\subsection{Populations with different co-operation pressures}
We first study the case of equally large populations, $\alpha_1=\alpha_2=1$, and focus on the effects of the co-operation pressures $u_1$ and $u_2$ on the stability and behaviour of the system. The resulting phase diagram has been reported in \cite{epl}, and is re-iterated in Fig. \ref{fig:pg} for completeness. The replicator dynamics, for any given realisation of the coupling matrices, evolve into a unique stable fixed point at large co-operation pressure. Below the phase transition lines depicted in Fig. \ref{fig:pg}, the system becomes non-ergodic in the sense that multiple microscopic stationary states exists, and initial conditions determine which of these is assumed asymptotically. For fully symmetric couplings $\Gamma=1$ the system still evolves into a fixed point, but exponentially many (in $N$) fixed points can be expected to exist \cite{DO,OD,OD2}. For $\Gamma<1$, i.e. for systems with partially or fully uncorrelated or anti-correlated couplings, no such behaviour is observed, and the dynamics may evolve towards a volatile, potentially chaotic state.

The role of the two co-operation pressures is to drive the respective populations into the internal region of their concentration simplices (defined by the constraints \\$N_1^{-1}\sum_{i=1}^{N_1}x_i=1$ and $N_2^{-1}\sum_j y_j=1$). Hence co-operation pressure promotes the survival of many pure strategies, or in other words a diverse stationary state. In order to study the effects of $u_1$ and $u_2$ we report results for a fixed value $u_1=1$ in Fig. \ref{fig:u1u2} and depict results for the diversity parameters $1/q_1$ and $1/q_2$ of the respective populations, as $u_2$ is varied. As expected both diversities are increasing functions of $u_2$. The effect on population $2$ is here much more direct, but remarkably, for small values of $u_2$ the diversity of population $1$ is a steep function of $u_2$ as well, at least for symmetric and asymmetric couplings. This indicates a feedback from population $2$ onto population $1$: decreasing the co-operation pressure on population $2$ leads to depletion in this population and this may then impact on the diversity of the other population as well, even though $u_1$ is kept constant.    

\begin{figure}
\vspace{0.75cm}
\resizebox{0.38\textwidth}{!}{%
  \includegraphics{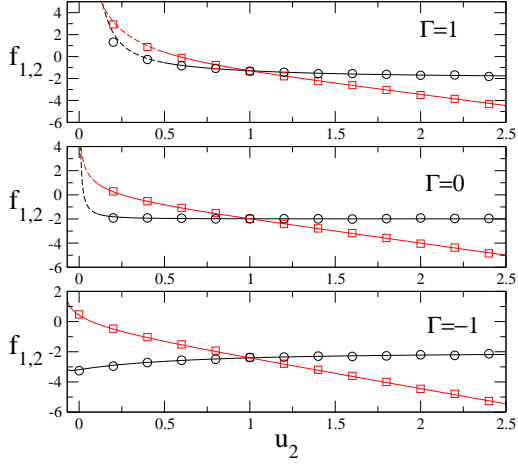}
}
\vspace*{4mm} \caption{(Colour on-line) Lagrange multipliers $f_1$ and $f_2$ of the two populations as a function of the co-operation pressure in population $2$. $u_1=1$ fixed ($\alpha_1=\alpha_2=1, c_1=c_2=0$). Simulation parameters, lines and symbols as in Fig. \ref{fig:u1u2}. Circles are population $1$, squares population $2$.}
\label{fig:u1u2fit}
\end{figure}

\begin{figure}
\vspace{0.75cm}
\resizebox{0.38\textwidth}{!}{%
  \includegraphics{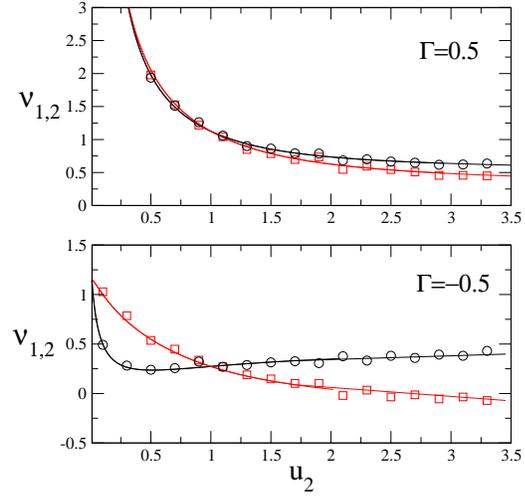}
}
\vspace*{4mm} \caption{(Colour on-line) Mean fitness $\nu_1$ and $\nu_2$ of the two populations as a function of the co-operation pressure in population $2$. $u_1=1$ fixed ($\alpha_1=\alpha_2=1,c_1=c_2=0$). Simulation parameters, lines and symbols as in Fig. \ref{fig:u1u2}, but $\Gamma=-0.5,0.5$ are chosen here. Symbols are from simulations, $N=100$ ($5000$ iteration steps, $20$ samples) with circles marking the fitness of population $1$, squares the one of population $2$.}
\label{fig:redfit}
\end{figure}

In Figs. \ref{fig:u1u2fit} and \ref{fig:redfit} we report on the fitness of the two populations. One here has to distinguish between the contributions of co-operation pressure and of the direct interaction to the overall fitness. While the Lagrange parameters $f_{1,2}$ entail the effects of co-operation pressures as well, we study the fitnesses $\nu_1=N_1^{-1}\sum_{ij}\overline{x_ia_{ij}y_j}$ and $\nu_2=N_2^{-1}\sum_{ji}\overline{y_jb_{ji}x_i}$ purely due to direct interaction in Fig. \ref{fig:redfit}. As shown in Fig. \ref{fig:u1u2fit} an increase of $u_2$ typically reduces $f_1$ and $f_2$, provided $\Gamma\ge 0$, i.e. provided that there is no degree of anti-correlation present in the coupling matrices. Again the effects of a change in $u_2$ are much more pronounced in $P2$ than in $P1$, and $f_1$ remains almost constant except for a region at small values of $u_2$. For negative values of $\Gamma$ the effect is quite different, as seen in the lower panel of Fig. \ref{fig:u1u2fit}. While, as before, $f_2$ is a decreasing function of $u_2$, non-trivial behaviour is induced in population $1$, and $f_1$ is found to be increasing in $u_2$, especially when $u_2$ small. This is the regime in which $u_2$ strongly controls the diversity of population $2$, and a reduction of the diversity in $P2$ appears to impact negatively on the fitness of $P1$. 

In order to disentangle the effects of co-operation pressure related contributions to the fitnesses, we study the behaviour of $\nu_n=f_n+2u_nq_n$, $n=1,2$ in Fig. \ref{fig:redfit}. For positive strategy correlation, we find that both fitnesses decrease as $u_2$ in increased. The non-trivial behaviour at negative strategy correlation remains, however, and $\nu_1$ is found to be a non-monotonous function of $u_2$, again signalling an indirect impact of the diversity of $P2$ on the payoff of $P1$. 

\begin{figure}
\vspace{0.75cm}
\resizebox{0.3\textwidth}{!}{%
  \includegraphics{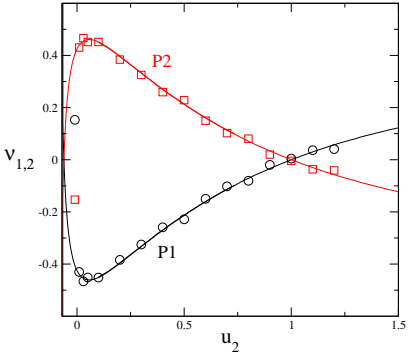}
}
\vspace*{4mm} \caption{(Colour on-line) Mean fitness $\nu_1$ and $\nu_2$ of the two populations as a function of the co-operation pressure in population $2$. $u_1=1$ fixed. Simulation parameters, lines and symbols as in Fig. \ref{fig:u1u2}, but $\Gamma=-1$. Symbols are from simulations, $N=100$ ($5000$ iteration steps, $200$ samples) with circles marking the fitness of population $1$, squares the one of population $2$.}
\label{fig:boing}
\end{figure}

An interesting effect is observed for the case $\Gamma=-1$, i.e. for
full anti-correlation, see Fig. \ref{fig:boing}. In order to
characterise the behaviour of $\nu_{1,2}$, we have here extended the
range of the co-operation pressures to include small negative values
of $u_2$. In the zero-sum case $\Gamma=-1$ one has $\nu_1=-\nu_2$ by
construction, and both fitnesses are found to be non-monotonic
functions of $u_2$ at fixed co-operation pressure in population
$1$. In particular a maximum of the fitness of $P2$ is found as the
co-operation pressure $u_2$ is increased, whereas the other population
experiences a minimum in fitness under these conditions.\footnote{We
here remark that while numerical experiments agree near perfectly with
the theoretical predictions for positive $u_2$, they become less
reliable at $u_2\approx 0$, where the theoretical curves are fairly
steep. The accuracy of simulations might here suffer from a small
number of surviving species and the associated finite-size effects and
sample-to-sample fluctuations.}

\subsection{Effects of relative population size}
\begin{figure}
\vspace{0.75cm}
\resizebox{0.38\textwidth}{!}{%
  \includegraphics{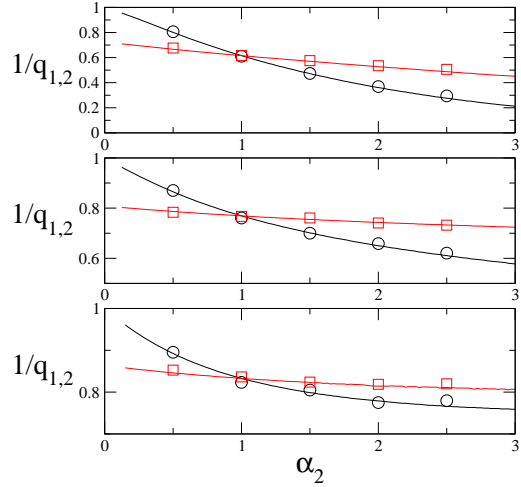}
}
\vspace*{4mm} \caption{(Colour on-line). Diversity parameters $1/q_1$ (circles) and $1/q_2$ (squares) as the size $\alpha_2$ of $P2$ is varied ($\alpha_1=1$ remains fixed; $u_1=u_2=1,c_1=c_2=0$). Simulations are for $N=200$, $20000$ integration steps, $10$ samples, solid lines are from the theory. $\Gamma=1,0,-1$ from top to bottom.}
\label{fig:a1a2}
\end{figure}

\begin{figure}
\vspace{0.75cm}
\resizebox{0.38\textwidth}{!}{%
  \includegraphics{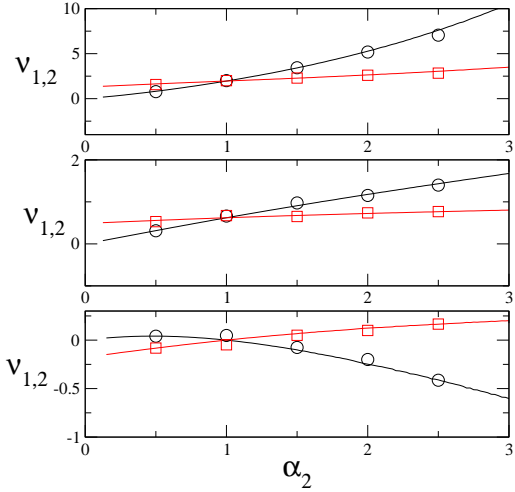}
}
\vspace*{4mm} \caption{(Colour on-line). Fitnesses $\nu_1$ (circles) and $\nu_2$ (squares) of the two populations, as the size $\alpha_2$ of $P2$ is varied ($\alpha_1=1$ remains fixed; $u_1=u_2=1$). Simulations are for $N=200$, $20000$ integration steps, $10$ samples, solid lines are from the theory. $\Gamma=1,0,-1$ from top to bottom.}
\label{fig:a1a2fit}
\end{figure}

The effects of varying the relative system size of the two populations
are illustrated in Figs. \ref{fig:a1a2} and \ref{fig:a1a2fit}. We here
first follow \cite{berg,berg2} and fix the size parameter $\alpha_1$
in the first population, and vary the size of $P2$. As shown in
Fig. \ref{fig:a1a2}, the diversities of both populations decrease as $\alpha_2=N_2/N$ is increased. The effect is much stronger in $P1$ than in $P2$. In
Fig. \ref{fig:a1a2fit} we depict the fitnesses $\nu_1,\nu_2$ of the
the two populations for this case of rectangular payoff matrices
$\alpha_1\neq\alpha_2$. We find that, within our settings and for
$\Gamma\geq 0$, both populations profit from reducing the aspect ratio
$\alpha_1/\alpha_2$, where the increase of fitness appears more
pronounced for $P1$ than for $P2$. Thus it appears it is the size of
the population with which a given species interacts, rather than the
size of its own population, which determines the fitness of this given
species. Inspection of the case of full anti-correlation (lower panel
in Fig. \ref{fig:a1a2fit}) shows that the fitness of $P1$ may display
non-monotonic behaviour as a function of $\alpha_2$. We note here that
$\nu_1/\nu_2=\alpha_2$ for the case of fully symmetric couplings
($\Gamma=1$), and that $\nu_1/\nu_2=-\alpha_2$ for $\Gamma=-1$,
i.e. for $a_{ij}=-b_{ji}$. These relations are due to the construction
of the model, where one has $\nu_1=N_1^{-1}\sum_{ij} \overline{x_i
a_{ij} y_j}$ and $\nu_2=N_2^{-1}\sum_{ji} \overline{y_j b_{ji} x_i}$
as well as $N_1/N_2=\alpha_1/\alpha_2$. No such simple relations
between the resulting $\nu_1,\nu_2$ are found for $-1<\Gamma<1$, as
$a_{ij}$ and $b_{ji}$ are then neither fully correlated nor fully
anti-correlated.

\begin{figure}
\vspace{0.75cm}
\resizebox{0.38\textwidth}{!}{%
  \includegraphics{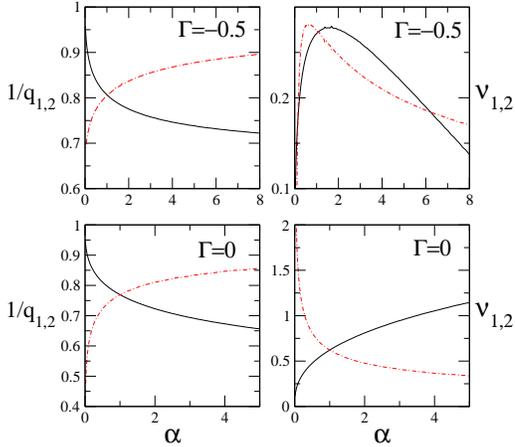}
}
\vspace*{4mm} \caption{(Colour on-line). Diversities and fitnesses $\nu_1$ of the two populations as a function of the aspect ratio $\alpha$ ($\alpha_1=\alpha^{1/2}, \alpha_2=\alpha^{-1/2}$). Upper panels are for $\Gamma=-0.5$, lower panels for $\Gamma=0$ ($u_1=u_2=1$). Solid lines are $P1$, dot-dashed lines $P2$.}
\label{fig:aspect}
\end{figure}

The results described so far in this section are not found to be
symmetrical around $\alpha_2=1$, In a symmetrical situation, where
only the aspect ratio $\alpha_1/\alpha_2$ matters, one would expect
that choosing $\alpha_2=a$ is equivalent to setting $\alpha_2=a^{-1}$
for any $a>0$, subject to a relabelling of the two
populations. Recall, however that the coupling strengths
$\overline{a_{ij}^2}=\overline{b_{ji}^2}$ remain fixed and are given
by $1/N$. The respective sizes of the two populations are
$N_1=\alpha_1 N$ and $N_2=\alpha_2 N$, where we fixed $N_1=N$
above. These definitions follow those of \cite{berg,berg2}. A more
symmetrical setup can be achieved by setting
e.g. $\alpha_1=\sqrt{\alpha}$ and $\alpha_2=1/\sqrt{\alpha}$, with
$0<\alpha<1$ the aspect ratio $\alpha=\alpha_1/\alpha_2=N_1/N_2$. As
seen in Fig. \ref{fig:aspect}, the diversity of population $1$
decreases as its relative size $\alpha$ is increases, whereas $1/q_2$
is monotonically decreasing (we do not report simulation results here
to keep the figure readable). The behaviour of the fitnesses crucially
depends on the symmetry of couplings, with multiple crossings of
$\nu_1$ and $\nu_2$ observed for suitable model parameters as shown in
Fig. \ref{fig:aspect}.

\subsection{Strategy correlations}

\begin{figure}
\vspace{0.75cm}
\resizebox{0.3\textwidth}{!}{%
  \includegraphics{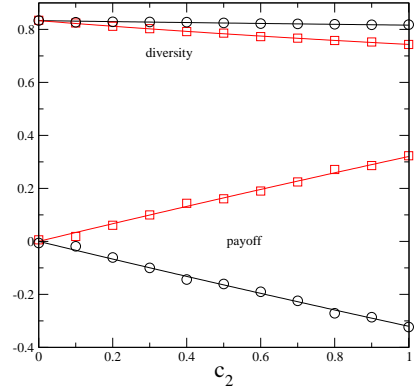}
}
\vspace*{4mm} \caption{(Colour on-line). Effects of strategy correlations. $c_1$ is fixed to $c_1=0$ as $c_2$ is varied from $0$ to $1$. Fully anti-symmetric (zero-sum) case $\Gamma=-1$. Co-operation pressures are $u_1=u_2=1$. Results from the fixed-point ansatz are shown as solid lines, markers are from simulations, with circles corresponding to $P1$ and squares to $P2$ ($N=100$, $10000$ integration steps, $200$ samples).}
\label{fig:c1c2}
\end{figure}

Finally, we have examined the effects of strategy correlations. We
here fix $\alpha_1=\alpha_2=1$ and $\Gamma=-1$. Results are presented
in Fig. \ref{fig:c1c2}, where we depict the case of varying $c_2$ at
fixed $c_1=0$. As seen in the figure, the diversity (or equivalently
the fraction of pure strategies played with non-zero probability)
decreases in both populations as $c_2$ is increased. The effect is
stronger in $P2$ than in $P1$. At the same time, the payoff for
population two increases as the correlation parameter $c_2$ is
increased, whereas the payoff for $P1$ decreases (note that
$\nu_1=-\nu_2$ by construction, as we are considering the case of
zero-sum games, $\Gamma=-1$). The condensation of $P2$ into a small
subset of strategies can here be understood as follows: an increased
value of $c_2$ induces correlations within the columns of the payoff
matrix $A=(a_{ij})$, so that in the extreme case of large $c_2\gg 1$,
the $\{a_{ij}\}_{j=1,\dots,N}$ become mostly independent of $i$. Thus
the payoff for both players becomes more and more independent of the
action of player $X$, and depends only on the choice $j$ of player
$Y$. In other words, the strategies available to player $X$ become
more and more alike as $c_2$ is increased, thus rendering some of
$Y$'s strategies generally beneficial for $Y$ (independently of the
choice of $X$), and others detrimental. Hence, $Y$ will mostly play
strategies beneficial from his perspective and will reduce the
diversity $1/q_2$ of his actions, while at the same time increasing
his payoff $\nu_2$, as seen in Fig. \ref{fig:c1c2}. See also \cite{berg} for similar cases. A more symmetrical
situation can be constructed by considering $c_1=c_2=c$, a case in
which correlations are present both in the rows and columns of the
payoff matrix. We do not depict results here, but only remark that in
this case the diversity of both populations decreases as $c$ is
increased (the payoff for both populations remains strictly zero in
this case due to the zero-sum property of the games considered, and
the exchange symmetry between $P1$ and $P2$).

\subsection{One-population models with heterogeneous co-operation pressure}
In this section we will consider a single-population replicator model with heterogeneous (i.e. species-dependent) co-operation pressures. Specifically consider $N$ species subject to the replicator equations
\be
\dot x_i = x_i\left(-2u_i x_i+\sum_{j=1}^N w_{ij} x_j-f\right), i=1,\dots,N
\ee
where the coupling matrix is a Gaussian random quantity as in \cite{OD} ($\overline{w_{ij}}=0, \overline{w_{ij}^2}=1/N, \overline{w_{ij}w_{ji}}=\Gamma/N)$, but where the co-operation pressure now carries a species index $i$, and where each $u_i$ is assumed to be drawn independently from the distribution $\rho(u)$ at the beginning of the dynamics and then remains fixed. Note that while the overall concentration $N^{-1}\sum_ix_i=1$ is conserved, species subject to a certain (e.g. high) co-operation pressure may well die out or be reduced in concentration to the advantage of species of a lesser co-operation pressure.

The further analysis leads to an ensemble of effective processes, one for each value of $u$ in the support of $\rho(u)$:
\BE
\dot x_u &=& x_u\bigg[-2u x_u -\Gamma\int dt' G(t,t')x_u(t')+\eta_u(t)-f(t)\bigg].\nonumber \\
\EE
The response function is now defined as \\ $G(t,t')=\int du \rho(u) \avg{\delta x_u(t)/\delta\eta_u(t')}$. Furthermore we have $\avg{\eta_u(t)\eta_u(t')}=\int du' \rho(u')\avg{x_{u'}(t)x_{u'}(t')}$ independently of $u$. A fixed point ansatz results in the following self-consistent equations for the integrated response $\chi$, the (inverse) diversity parameter $q$ and the fitness $f$:
\BE
\chi&=&g_0(\Delta)\int du \frac{\rho(u)}{(2u-\Gamma\chi)},\nonumber \\
q^{-1/2}&=&g_1(\Delta)\int du \frac{\rho(u)}{(2u-\Gamma\chi)},\nonumber\\
1&=&g_2(\Delta)\int du \frac{\rho(u)}{(2u-\Gamma\chi)^2},\label{eq:het2}
\EE
where $\Delta=-f/\sqrt{q}$. While we note that the fitnesses $f$ of any
surviving species come out as equal (and independent of their
co-operation pressures), their relative concentrations
\be
C(u)=\frac{1}{|I(u)|}\sum_{i\in I(u)} x_i
\ee
and second moments
\be
Q(u)=\frac{1}{|I(u)|}\sum_{i\in I(u)} x_i^2
\ee
(suitable sample-averages are implied) maybe well be dependent on $u$ \footnote{We here note that the fraction of survivors $\phi(u)=\sum_{i\in I(u)}\Theta(x_i)$ (with $\Theta(\cdot)$ the step function) comes out as independent of $u$, and is given by $\phi(u)\equiv \phi=g_0(\Delta)$.}. Here $I(u)$ denotes the set of species $i\in\{1,\dots,N\}$ with co-operation pressure $u_i\in[u-du,u+du]$, with $du$ an infinitesimal element. Overall self-consistency requires that $\int du \rho(u)C(u)=1$ and $\int du \rho(u)Q(u)=q$.

\begin{figure}
\vspace{1.25cm}
\resizebox{0.38\textwidth}{!}{%
  \includegraphics{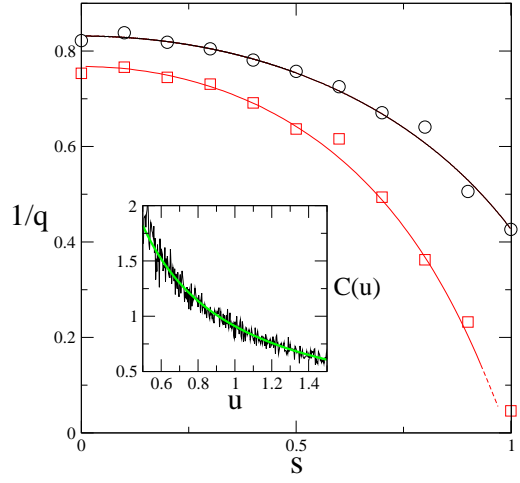}
}
\vspace*{4mm} \caption{(Colour on-line) One-population model with heterogeneous co-operation pressures. Main panel: diversity versus variability of co-operation pressure ($\mu=1$ fixed). Upper curve shows $\Gamma=-1$, lower curve $\Gamma=0$. Symbols from simulations ($N=300$, $5000$ integration steps, $10$ samples). Inset: concentration of sub-population with co-operation pressure $u$ as a function of $u$ ($\Gamma=0, \mu=1, s=0.5$). Solid line is theory, noisy line simulations ($N=300$, $100$ samples). Dashed line to the lower right in main panel indicates unstable regime for $\Gamma=0$ }
\label{fig:het}
\end{figure}

We will here restrict to the case of a flat distribution $\rho(u)$ over the interval $[\mu-s,\mu+s]$, so that the integrals on the right-hand side of (\ref{eq:het2}) can be performed explicitly to give
\BE
\chi&=&g_0(\Delta)(4s)^{-1}\ln\left[\frac{2\mu+2s-\Gamma\chi}{2\mu-2s-\Gamma\chi}\right],\nonumber \\
q^{-1/2}&=&g_1(\Delta)(4s)^{-1}\ln\left[\frac{2\mu+2s-\Gamma\chi}{2\mu-2s-\Gamma\chi}\right],  \nonumber\\
1&=&\frac{g_2(\Delta)}{(2\mu-\Gamma\chi)^2-4s^2}.
\EE
Results are shown and compared with simulations in Fig. \ref{fig:het}. Interestingly one observes a decline in diversity of the population as the variability $s$ of the co-operation pressure is increased. Thus increasing the complexity of the properties of the individual species might lead to a less diverse composition of the eco-system at stationarity. This effect seems to be mostly independent of the symmetry of the couplings\footnote{We have not been able to solve the resulting fixed-point equations for $\Gamma=1$ at large widths $s$ of the distribution of co-operation pressures.}. The inset of Fig. \ref{fig:het} demonstrates that the relative weight $C(u)$ of species subject to co-operation $u$ decreases non-linearly with $u$, so that species with comparably low co-operation pressure dominate the population at the fixed point. More precisely $C(u)$ is found to be given by $C(u)=\sqrt{q}g_1(\Delta)/(2u-\Gamma\chi)$, i.e. it roughly decays as the inverse power of $u$. As seen in the inset of the figure the theoretical prediction of this behaviour agrees perfectly with results from simulations.

\section{Conclusions}
In summary, we have extended existing generating functional techniques to study the behaviour of two-population replicator systems, and have focused on the effects of co-operation pressure, relative population size and strategy correlation. A phase transition between a stable phase with one unique fixed point of the replicator dynamics has been identified, and characterised analytically. This phase is separated from a second unstable phase by a transition line in parameter space, which can be determined from the statistical mechanics analysis. Our study demonstrates that control parameters such as co-operation pressure, aspect ratio and strategy correlation have non-trivial effects on the the system of replicators, and can induce non-monotonic behaviour of the resulting fitnesses. We have also addressed single population models with species-dependent co-operation pressure. The statistical mechanics theory is then formulated in terms of an ensemble of effective processes, one for each value of the co-operation pressure present in the population. We find that variability in the co-operation pressures reduces the diversity of the set of surviving species. In such a mixed population, the weight of species subject to a specific co-operation pressure $u$ scales as $u^{-1}$ asymptotically.

Natural extensions of the present model include the generalisation to
a larger number of populations, and the study of dynamics which is
different from standard replicator equations \cite{Book4,Book5,traulsen}. Furthermore,
relatively little is known about the non-ergodic phase of random
replicator systems, so that future work might address the properties
of such phases. From the point of view of real-world eco-systems and
population dynamics the assumption of a fully connected interaction
matrix is at best a crude approximation, and would ideally need to be
replaced by ensembles with sparse interactions. While the analysis of
fully connected replicator models like the one discussed in this
paper, is relatively straightforward and leads to an effective theory
in terms of two-time quantities (the correlation and response
functions), dilute replicator systems pose a much more demanding
challenge, as order parameter equations do not close on the two-time
level \cite{hatchett}. Future work might hence address such models,
potentially relying on recently developed techniques to study spin
systems with sparse interaction matrices \cite{weigt}.
\section*{Acknowledgements}

This work was supported through an RCUK Fellowship (University of Manchester, RCUK reference EP/E500048/1), and by EU NEST No. 516446, COMPLEXMARKETS (ICTP Trieste).

\end{document}